\documentstyle[preprint,aps,epsfig]{revtex}
\tightenlines
\begin{document}
%
%
%
%
%
\draft
\title{
%
%
%
%
$I=2$ Pion Scattering Length with the Wilson Fermion
}
%
%
\author{
       S.~Aoki,$^{\rm 1    }$
   M.~Fukugita,$^{\rm 2    }$
  S.~Hashimoto,$^{\rm 3    }$
 K-I.~Ishikawa,$^{\rm 1, 4 }$
   N.~Ishizuka,$^{\rm 1, 4 }$
    Y.~Iwasaki,$^{\rm 4    }$
     K.~Kanaya,$^{\rm 1, 4 }$
     T.~Kaneko,$^{\rm 3    }$
  Y.~Kuramashi,$^{\rm 3    }$
      M.~Okawa,$^{\rm 5    }$
      T.~Onogi,$^{\rm 6    }$
   S.~Tominaga,$^{\rm 4    }$
    N.~Tsutsui,$^{\rm 3    }$
      A.~Ukawa,$^{\rm 1, 4 }$
     N.~Yamada,$^{\rm 3    }$
 T.~Yoshi\'{e},$^{\rm 1, 4 }$ \\
(JLQCD Collaboration)
}
\address{
${}^{\rm 1}$ Institute of Physics, University of Tsukuba            , Tsukuba, Ibaraki   305-8571, Japan \\
${}^{\rm 2}$ Institute for Cosmic Ray Research, University of Tokyo , Kashiwa, Chiba     277-8582, Japan \\
${}^{\rm 3}$ High Energy Accelerator Research Organization (KEK)    , Tsukuba, Ibaraki   305-0801, Japan \\
${}^{\rm 4}$ Center for Computational Physics, University of Tsukuba, Tsukuba, Ibaraki   305-8577, Japan \\
${}^{\rm 5}$ Department of Physics, Hiroshima University, Higashi-Hiroshima  , Hiroshima 739-8526, Japan \\
${}^{\rm 6}$ Yukawa Institute for Theoretical Physics, Kyoto University      , Kyoto     606-8502, Japan \\
}
\date{
\today
}
\maketitle
%
%
\setlength{\baselineskip}{18pt}
%
%
\begin{abstract}
The calculation of the $I=2$ pion scattering length in quenched lattice QCD
is revisited.
The calculation is carried out with the Wilson fermion action 
employing L\"uscher's finite size scaling method 
at $\beta=5.9$, $6.1$, and $6.3$ corresponding 
to the range of 
lattice spacing $a\simeq 0.12 - 0.07$~fm.
We obtain in the continuum limit 
$a_0/m_\pi = - 2.09(35)\ 1/{\rm GeV}^2$,
which is consistent
with the prediction of chiral perturbation theory 
$a_0/m_\pi = - 2.265(51)\ 1/{\rm GeV}^2$. 
\end{abstract}
%
%
%
%
\narrowtext
%
%
Lattice calculations of $S$-wave scattering lengths of the two-pion system
are important step to understand dynamical effects of strong interactions. 
There are already a number of calculations for the $I=2$ process
with either Staggered~\cite{Kuramashi,SGK} or Wilson fermion 
action~\cite{Kuramashi,GPS}.
While these calculations gave results that are in gross agreement
with the prediction of chiral perturbation theory (CHPT)~\cite{CHPT_a0},
they were made on coarse and small lattices. More importantly,
the continuum extrapolation was not made.
Aiming to improve on these points, 
we carried out a calculation of the $I=2$ $S$-wave scattering 
length in quenched lattice QCD. A preliminary result was
reported in Ref.~\cite{JLQCD}, in which some
disagreement with the CHPT prediction was mentioned.
In the mean time Liu {\it et.al.} carried out a similar calculation 
with the improved gauge and the Wilson fermion actions
on anisotropic lattices~\cite{LZCM}.

We employ the standard plaquette action for gluons and the Wilson action 
for quarks, and explore the parameter range 
$m_\pi / m_\rho \sim 0.55 - 0.80$ for the chiral extrapolation and 
$a \sim 0.07 - 0.12$~fm for the continuum extrapolation.
This is compared with the parameters of Liu {\it et.al.} which range
$m_\pi / m_\rho \sim 0.7 - 0.9$ and $a_s \sim 0.2 - 0.4$~fm. 
Our calculations are made for parameters significantly closer 
to the chiral limit. 
In this short report we give the final result of our analysis. 

The numbers of configurations (lattice sizes) are
$187~ (16^3 \times 64 )$,
$120~ (24^3 \times 64 )$, and
$100~ (32^3 \times 80 )$ for $\beta=5.9$, $6.1$, and $6.3$, respectively.
Quark propagators are solved with the Dirichlet boundary condition
in the time direction
and the periodic boundary condition in the space directions.
The pion mass covers the range of $450 - 900{\rm MeV}$.
The lattice constant is estimated from the $\rho$ meson mass, 
which was obtained in our previous study\cite{JLQCDfB}, to be 
$a^{-1} = 1.64(2), 2.29(4), 3.02(5)$~(GeV) at 
$\beta=5.9$, $6.1$, and $6.3$.
Our calculations were carried out on the Fujitsu VPP500/80 
supercomputer at KEK.

The energy eigenvalue of a two-pion system in a finite periodic box $L^3$
is shifted by the finite-size effect.
L\"uscher presented a relation between the energy shift $\Delta E$
and the $S$-wave scattering length $a_0$, given by~\cite{Luscher}
\begin{equation}
   - \Delta E \cdot  \frac{ m_\pi L^2 }{ 4 \pi^2 }
= T + C_1 \cdot T^2 + C_2 \cdot T^3 + O( T^4 ) \ ,
\label{Luscher.eq}
\end{equation}
where $T = a_0 / ( \pi L )$. The constants are 
$C_1 = -8.9136$ and $C_2 = 62.9205$ computed from  geometry of the lattice.
Since $T$ has a small value, typically $\sim - 10^{-2}$ in our simulation,
we can safely neglect the higher order terms $O(T^4)$.

The energy shift $\Delta E$ can be obtained from the ratio
$R( t ) = G( t )/D( t )$, where
\begin{eqnarray}
&&   G( t ) = \langle \pi^+(t) \pi^+ (t) W^{-}(t_1) W^{-}(t_2 ) \rangle  \cr
&&   D( t ) = \langle \pi^+(t) W^{-} (t_1)  \rangle \
              \langle \pi^+(t) W^{-} (t_2)  \rangle \ .
\label{eq:GD}
\end{eqnarray}
In order to enhance signals against the noise 
we use wall sources for $\pi^-$, which are denoted by $W^{-}$ in (\ref{eq:GD}),
by fixing gauge configurations to the Coulomb gauge.
The two wall sources are placed at different time slices $t_1$ and $t_2$
to avoid contaminations from Fierz-rearranged terms in the two-pion state
which would occur for $t_1=t_2$.
We set $t_2=t_1+1$ and
$t_1=8$, $10$, $13$ for $\beta=5.9$, $6.1$, $6.3$.

\begin{figure}
\begin{center}
    \epsfig{ file=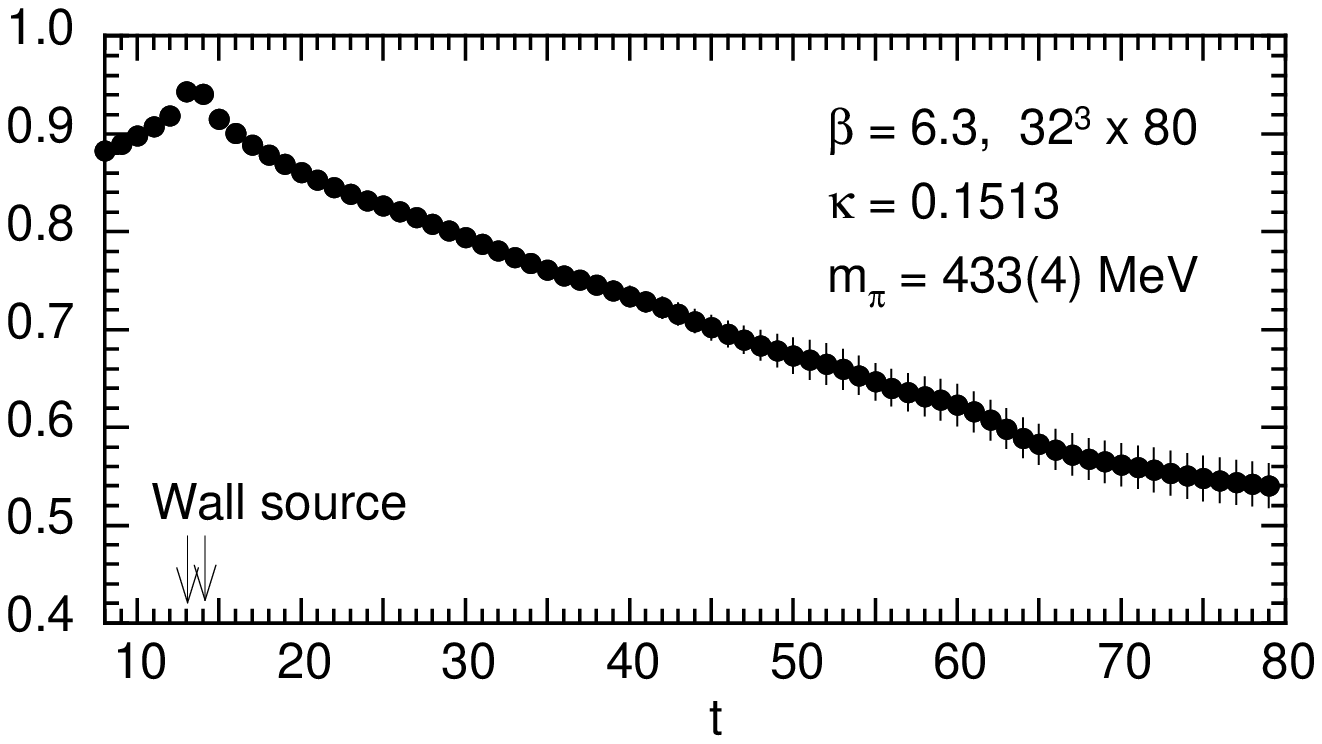, width=11.0cm }
\end{center}
\caption{ \label{TR.63.00.fig}
The ratio $R(t)=G(t)/D(t)$ at $\beta=6.3$ and $\kappa=0.1513$
corresponding to $m_\pi=433(4){\rm MeV}$.
The wall sources are located at $t=13$ and $14$.
}
\end{figure}

An example of $R(t)$ is plotted in Fig.~\ref{TR.63.00.fig}
for $\beta=6.3$ and $\kappa =0.1513$
corresponding to $m_\pi= 433(4){\rm MeV}$.
We see a clear, almost linear fall-off as a function of $t$ till $t=80$
even for a small energy $\Delta E\approx 20$ MeV, showing that 
our wall sources work well for the two-pion state.

The energy shift $\Delta E$ is obtained from the linear 
term in the expansion of $R(t)$ :
\begin{equation}
R(t) = Z \cdot \biggl( 
           1 - \Delta E \cdot \tau   +   {\rm O}( \tau^2 )  \biggr)
\ ,
\label{GB_eq}
\end{equation}
where $\tau = t - t_2$.
The quadratic and higher order terms have no simple relations to $\Delta E$
due to effects from intermediate off-shell two-pion states~\cite{SGK} and 
quenching effects~\cite{Bernard-Golterman}.
We first attempt to fit the data with the form
\begin{equation}
{\bf (Sqr)}  \quad Z \cdot \biggl( 1 - \Delta E \cdot \tau +  E' \cdot \tau^2  \biggr)
\ .
\label{eq:sqr}
\end{equation}
We find that this fit ${\bf (Sqr)}$ is quite ill-determined, since
the two terms correlate so strongly, resulting in unacceptably
large errors in $\Delta E$ and $E'$.  
We then attempt to fit with 
\begin{eqnarray}
&& {\bf (Exp)}    \quad Z \cdot \exp ( - \Delta E \cdot \tau )            \ , \cr
&& {\bf (Lin)} \  \quad Z \cdot \biggl( 1 - \Delta E \cdot \tau  \biggr)  \ .
\label{fit_R.eq}
\end{eqnarray}
These fitting forms give well-determined $\Delta E$, while it may be
contaminated by contributions from  the second order term.  
We also include a fit of the form 
\begin{equation}
{\bf (Old)}  \quad Z - \Delta E \cdot \tau  
\label{JLQCD.OLD_eq}
\end{equation}
into our attempts for completeness, since this was used in our
preliminary report~\cite{JLQCD}. Note, however, that this form is
theoretically correct only when $Z$ is close to unity.
The results for $\Delta E$ (and $E'$ in case ${\bf (Sqr)}$)
are given in 
Table~\ref{result.59.table} for $\beta=5.9$, 
Table~\ref{result.61.table} for $\beta=6.1$, and 
Table~\ref{result.63.table} for $\beta=6.3$.
We take the same fitting range for the four fits,
$t=21 - 42$ for $\beta=5.9$,
$t=25 - 50$ for $\beta=6.1$, and
$t=27 - 62$ for $\beta=6.3$.
The value of $\chi^2$ for each fitting is always small, and does not
discriminate among fits. We do not consider case ${\bf (Sqr)}$ further
because of very large errors, although the resulting 
central values for the energy 
shift are consistent with those from ${\bf (Exp)}$ and 
${\bf (Lin)}$. The problem we must consider is whether we can remove
contaminations of the second order term for $\Delta E$ from
${\bf (Exp)}$ and ${\bf (Lin)}$. 

\begin{figure}
\begin{center}
    \epsfig{ file=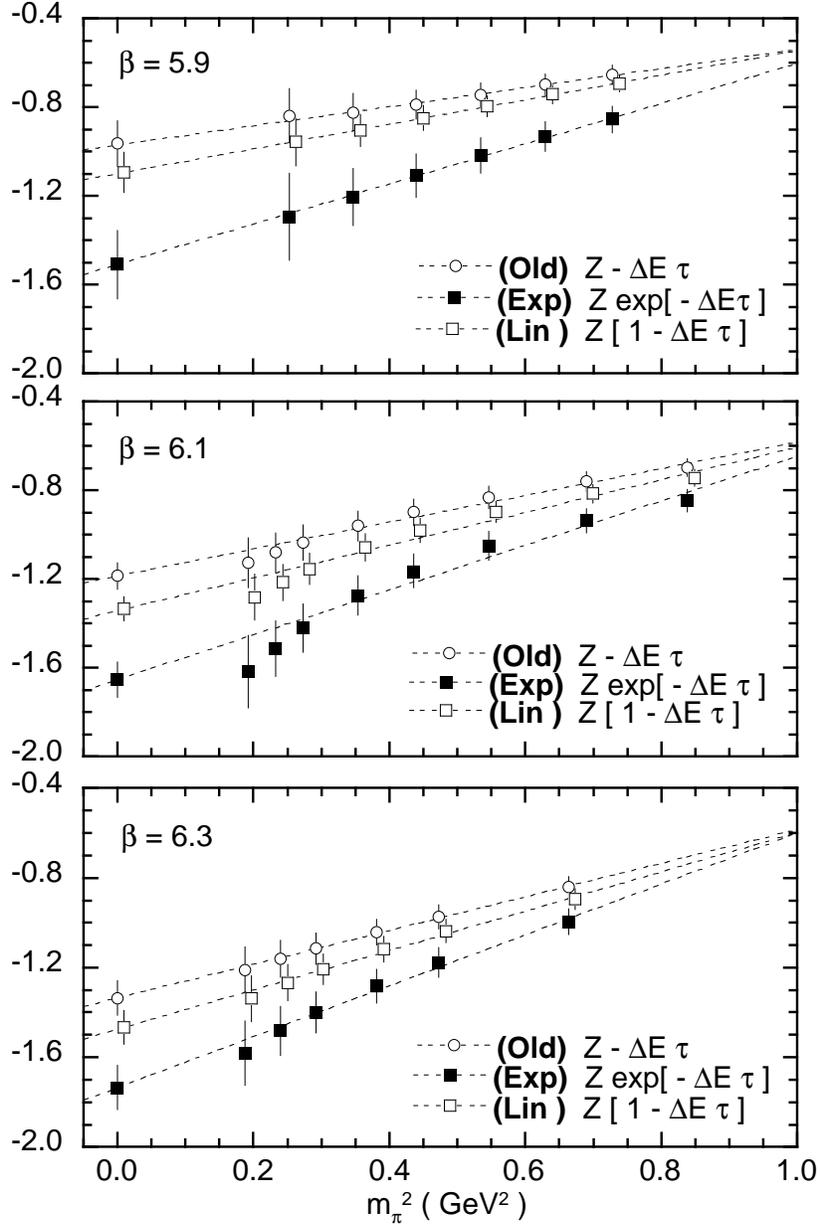, width=11.0cm }
\end{center}
\caption{ \label{a0-mp.fig}
The  mass dependence of 
$a_0 / m_{\pi} ( 1/{\rm GeV}^2 )$ at each lattice spacing.
}
\end{figure}

Figure~\ref{a0-mp.fig}
shows $a_0/m_\pi$ as a function of the pion mass obtained at each $\beta$, 
with their numerical values tabulated in
Table~\ref{result.59.table}, \ref{result.61.table}, and \ref{result.63.table}.
We observe a large difference between ${\bf (Exp)}$ and ${\bf (Lin)}$,
indicating that contributions from the $O(\tau^2)$ term are indeed non-negligible
and largely affect the determination of $\Delta E$.
The common in all figures of $a_0/m_\pi$ versus $m_\pi$ is that the data
show a behavior linear in $m_\pi^2$. 
We then fit 
\begin{equation}
  a_0/m_\pi = A + B \cdot m_\pi^2 
\label{chiral-fit.eq}
\end{equation}
to extract the value $A$ in the chiral limit. 
From the view point of CHPT
we may in principle have a term $m_\pi^2 \log( m_\pi^2 / \Lambda^2 )$
added to (\ref{chiral-fit.eq}). 
If we include this term with a free coefficient into the fit, 
however, the coefficients correlate too strongly that the fit is invalidated, 
producing a large error also for $A$. 
It is difficult to distinguish $m_\pi^2$ and 
$m_\pi^2 \log( m_\pi^2)$ within the range of $m_\pi^2$ 
that concerns us and the limited statistics. 
Since we do not see any significant curvature in the figure of 
$a_0/m_\pi$ versus $m_\pi$,
we simply drop this logarithmic term which itself vanishes at the
chiral limit. We also note that for the Wilson fermion action 
the term $\propto 1/m_\pi^2$ may also exist, arising from explicit 
breaking of chiral symmetry, and also from quenching 
effects~\cite{Bernard-Golterman}. We do not see a $1/m_\pi^2$ effect, 
as our simulation is perhaps well away from $m_\pi^2=0$ and such a 
term is already damped into noise for the range of our simulation. 
Hence we do not include this term into our fit. 
In order to detect these two additional terms 
a simulation is needed close to the chiral limit with much higher statistics.  

\begin{figure}
\begin{center}
    \epsfig{ file=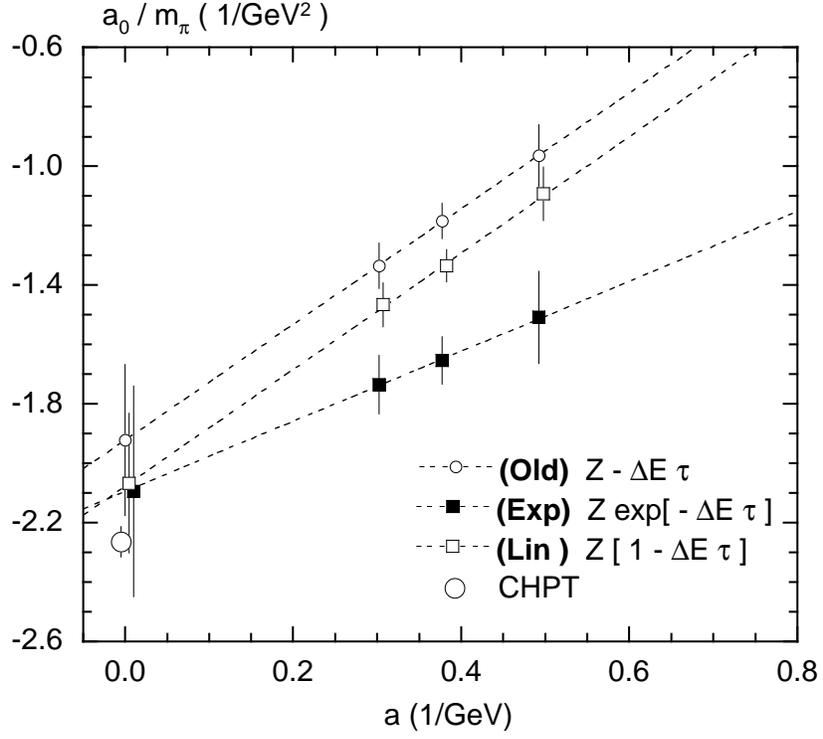, width=11.0cm }
\end{center}
\caption{ \label{a0-mp.C-limit.fig}
$a_0 / m_{\pi} ( 1/{\rm GeV}^2 )$ at the chiral limit at each lattice spacing.
The CHPT prediction 
is also plotted.
}
\end{figure}

In Fig.~\ref{a0-mp.C-limit.fig} we present $a_0 / m_\pi$ in the 
chiral limit as a function of the lattice spacing,
together with continuum extrapolations.
Their numerical values are tabulated in Table.~\ref{a0-mp.C-limit.table}
where values for ${\bf (Sqr)}$ are also listed for completeness.
This figure demonstrates a sizable scaling violation, but
exhibits a very clean linear dependence as a function of $a$.
It is interesting to observe that the difference between ${\bf (Exp)}$ and 
${\bf (Lin)}$, which are quite sizable on finite lattices, 
vanishes approaching the continuum limit.
This shows that the second order term
$O(\tau^2)$ included in (\ref{GB_eq})
becomes irrelevant as $\Delta E \cdot \tau$ 
becomes sufficiently small;
one may use any formula correct to the first order in $\tau$ 
to extract the $\Delta E$. 
On the other hand, the extrapolation with ${\bf (Old)}$ gives a value 
somewhat different from
the other two in the continuum limit, indicating that the departure of
$Z$ from unity could be non-negligible (although at $1.2-1.5 \sigma$).

As our final value for the scattering length in the continuum limit
at physical pion mass
we take the result from ${\bf (Exp)}$, which agrees with 
that from ${\bf (Lin)}$ but has a larger statistical error: 
\begin{equation}
   a_0 / m_\pi = - 2.09(35) \ 1/{\rm GeV}^2  \ , 
\end{equation}
where a rather large error arises from the continuum extrapolation.
This result is compared with the CHPT prediction : 
\begin{equation}
   a_0 / m_\pi = - 2.265(51)  \ 1/{\rm GeV}^2 \ ,
\end{equation}
The scattering length we derived at the continuum limit 
agrees well with the prediction of CHPT.
The difference seen in the fitting formula of ${\bf (Old)}$
and ${\bf (Lin)}$ accounts for the $1.5 \sigma$ difference of the lattice
result from the CHPT prediction mentioned in our preliminary report, 
which is based on an incorrect extrapolation formula ${\bf (Old)}$.

We remark that our results also agree with those of Liu {\it et.al.}~\cite{LZCM}
\begin{eqnarray}
&& a_0 / m_\pi = - 1.75(38) \ 1/{\rm GeV}^2 \quad \mbox{ for Scheme I  } \ , \\
%
&& a_0 / m_\pi = - 2.34(46) \ 1/{\rm GeV}^2 \quad \mbox{ for Scheme II } \ ,  
%
\end{eqnarray}
where two values (Scheme I and II) refer to
their two different treatments for the finite volume corrections.

In this article we have reported a calculation of 
the scattering length for the $I=2$ $S$-wave two pion system.
We have shown that the result in the continuum limit 
is virtually independent of the choice of fitting functions 
used to extract $\Delta E$ from the ratio $R(t)$, 
and that it is consistent with the prediction of CHPT
within our 15\% statistical error. 
%
%

%
%
%
%
%
%
%
%
\newpage
\begin{table}[t]
\begin{center}
\begin{tabular}{ l l l l l }
$\beta=5.9$   \\
$\kappa                   $ & Fit & $\Delta E        $ & $E'              $ & $a_0 / m_\pi    $ \\
$m_{\pi}^2 \ ({\rm GeV}^2)$ &     & $(\times 10^{-3})$ & $(\times 10^{-5})$ & $(1/{\rm GeV}^2)$ \\
\hline
$0.1585    $ & ${\bf Old} $ & $12.4(21) $ & $-      $ & $-0.84(12) $ \\
$0.2529(56)$ & ${\bf Exp} $ & $20.9(40) $ & $-      $ & $-1.29(20) $ \\
             & ${\bf Lin} $ & $14.5(19) $ & $-      $ & $-0.96(11) $ \\
             & ${\bf Sqr} $ & $23.1(74) $ & $29(21) $ & $-1.40(35) $ \\
\hline                                                          
$0.1580    $ & ${\bf Old} $ & $12.5(15) $ & $-      $ & $-0.822(84)$ \\
$0.3468(49)$ & ${\bf Exp} $ & $19.9(27) $ & $-      $ & $-1.20(13) $ \\
             & ${\bf Lin} $ & $14.0(13) $ & $-      $ & $-0.905(72)$ \\
             & ${\bf Sqr} $ & $19.0(57) $ & $14(15) $ & $-1.16(27) $ \\
\hline                                                    
$0.1575    $ & ${\bf Old} $ & $12.1(12) $ & $-      $ & $-0.786(65)$ \\
$0.4396(48)$ & ${\bf Exp} $ & $18.5(21) $ & $-      $ & $-1.108(98)$ \\
             & ${\bf Lin} $ & $13.3(11) $ & $-      $ & $-0.849(56)$ \\
             & ${\bf Sqr} $ & $16.3(50) $ & $ 8(13) $ & $-1.00(24) $ \\
\hline                                                    
$0.1570    $ & ${\bf Old} $ & $11.5(10) $ & $-      $ & $-0.743(55)$ \\
$0.5337(49)$ & ${\bf Exp} $ & $17.0(17) $ & $-      $ & $-1.017(79)$ \\
             & ${\bf Lin} $ & $12.48(92)$ & $-      $ & $-0.794(47)$ \\
             & ${\bf Sqr} $ & $14.4(45) $ & $ 5(12) $ & $-0.89(22) $ \\
\hline                                                  
$0.1565    $ & ${\bf Old} $ & $10.86(91)$ & $-      $ & $-0.698(48)$ \\
$0.6297(50)$ & ${\bf Exp} $ & $15.6(15) $ & $-      $ & $-0.931(67)$ \\
             & ${\bf Lin} $ & $11.69(82)$ & $-      $ & $-0.741(42)$ \\
             & ${\bf Sqr} $ & $13.0(41) $ & $ 3(10) $ & $-0.81(20) $ \\
\hline                                                  
$0.1560    $ & ${\bf Old} $ & $10.19(82)$ & $-      $ & $-0.654(43)$ \\
$0.7279(51)$ & ${\bf Exp} $ & $14.2(13) $ & $-      $ & $-0.855(59)$ \\
             & ${\bf Lin} $ & $10.92(75)$ & $-      $ & $-0.692(38)$ \\
             & ${\bf Sqr} $ & $11.9(37) $ & $2.6(95)$ & $-0.74(19) $ \\
\end{tabular}
\end{center}
\caption{\label{result.59.table}
The results at $\beta=5.9$. 
Four lines for each $m_\pi$ are results with the fitting functions 
${\bf Old}$, ${\bf Exp}$, ${\bf Lin}$, and ${\bf Sqr}$,
which are defined in (\ref{eq:sqr})--(\ref{JLQCD.OLD_eq}).
}
\end{table}
%
%
\begin{table}[t]
\begin{center}
\begin{tabular}{ l l l l l }
$\beta=6.1$   \\
$\kappa                $ & Fit & $\Delta E        $ & $E'              $ & $a_0 / m_\pi    $ \\
$m_\pi^2\ ({\rm GeV}^2)$ &     & $(\times 10^{-3})$ & $(\times 10^{-5})$ & $(1/{\rm GeV}^2)$ \\
\hline
$0.15430   $ & ${\bf Old} $ & $8.45(98) $ & $-      $ & $-1.13 (11)$ \\
$0.1925(42)$ & ${\bf Exp} $ & $13.0(17) $ & $-      $ & $-1.62(17) $ \\
             & ${\bf Lin} $ & $9.82(95) $ & $-      $ & $-1.28(10) $ \\
             & ${\bf Sqr} $ & $14.2(37) $ & $9.9(84)$ & $-1.73(36) $ \\
\hline                                                      
$0.15415   $ & ${\bf Old} $ & $8.17(79) $ & $-      $ & $-1.080(89)$ \\           
$0.2329(42)$ & ${\bf Exp} $ & $12.2(13) $ & $-      $ & $-1.51(13) $ \\           
             & ${\bf Lin} $ & $9.38(76) $ & $-      $ & $-1.214(82)$ \\           
             & ${\bf Sqr} $ & $13.0(33) $ & $8.2(74)$ & $-1.59(32) $ \\
\hline                                                      
$0.15400   $ & ${\bf Old} $ & $7.88(71) $ & $-      $ & $-1.035(79)$ \\           
$0.2732(42)$ & ${\bf Exp} $ & $11.6(11) $ & $-      $ & $-1.42(11) $ \\           
             & ${\bf Lin} $ & $8.97(68) $ & $-      $ & $-1.154(72)$ \\           
             & ${\bf Sqr} $ & $12.2(30) $ & $7.2(68)$ & $-1.48(29) $ \\
\hline                                                      
$0.15370   $ & ${\bf Old} $ & $7.38(62) $ & $-      $ & $-0.960(68)$ \\           
$0.3539(44)$ & ${\bf Exp} $ & $10.43(93)$ & $-      $ & $-1.274(89)$ \\           
             & ${\bf Lin} $ & $8.23(59) $ & $-      $ & $-1.056(62)$ \\           
             & ${\bf Sqr} $ & $11.0(26) $ & $6.2(60)$ & $-1.33(25) $ \\
\hline                                                      
$0.15340   $ & ${\bf Old} $ & $6.96(56) $ & $-      $ & $-0.987(60)$ \\          
$0.4355(46)$ & ${\bf Exp} $ & $9.56(80) $ & $-      $ & $-1.164(76)$ \\          
             & ${\bf Lin} $ & $7.73(53) $ & $-      $ & $-0.980(55)$ \\          
             & ${\bf Sqr} $ & $10.2(24) $ & $5.5(55)$ & $-1.22(22) $ \\
\hline                                                      
$0.15300   $ & ${\bf Old} $ & $6.48(49) $ & $-      $ & $-0.831(53)$ \\          
$0.5465(49)$ & ${\bf Exp} $ & $8.65(68) $ & $-      $ & $-1.050(65)$ \\          
             & ${\bf Lin} $ & $7.12(47) $ & $-      $ & $-0.898(48)$ \\          
             & ${\bf Sqr} $ & $9.3(21)  $ & $4.7(49)$ & $-1.11(20) $ \\
\hline                                                      
$0.15250   $ & ${\bf Old} $ & $5.96(43) $ & $-      $ & $-0.760(45)$ \\          
$0.6897(52)$ & ${\bf Exp} $ & $7.73(58) $ & $-      $ & $-0.938(56)$ \\          
             & ${\bf Lin} $ & $6.48(42) $ & $-      $ & $-0.814(42)$ \\          
             & ${\bf Sqr} $ & $8.2(19)  $ & $3.9(43)$ & $-0.99(18) $ \\
\hline                                                      
$0.15200   $ & ${\bf Old} $ & $5.48(40) $ & $-      $ & $-0.697(41)$ \\          
$0.8385(55)$ & ${\bf Exp} $ & $6.95(52) $ & $-      $ & $-0.845(50)$ \\          
             & ${\bf Lin} $ & $5.92(38) $ & $-      $ & $-0.743(39)$ \\          
             & ${\bf Sqr} $ & $7.4(17)  $ & $3.1(39)$ & $-0.89(16) $ \\
\end{tabular}
\end{center}
\caption{\label{result.61.table}
The results at $\beta=6.1$.
Four lines for each $m_\pi$ are results with the fitting functions 
${\bf Old}$, ${\bf Exp}$, ${\bf Lin}$, and ${\bf Sqr}$,
which are defined in (\ref{eq:sqr})--(\ref{JLQCD.OLD_eq}).
}
\end{table}
%
%
\begin{table}[t]
\begin{center}
\begin{tabular}{ l l l l l }
$\beta=6.3$   \\
$\kappa                  $ & Fit & $\Delta E        $ & $E'              $ & $a_0 / m_\pi    $ \\
$m_{\pi}^2\ ({\rm GeV}^2)$ &     & $(\times 10^{-3})$ & $(\times 10^{-5})$ & $(1/{\rm GeV}^2)$ \\
\hline
$0.15130   $ & ${\bf Old} $ & $5.97(60)$ & $-      $ & $-1.21(11) $ \\
$0.1876(36)$ & ${\bf Exp} $ & $8.19(89)$ & $-      $ & $-1.58(14) $ \\
             & ${\bf Lin} $ & $6.71(60)$ & $-      $ & $-1.34 (10)$ \\
             & ${\bf Sqr} $ & $7.9(18) $ & $2.4(36)$ & $-1.54(29) $ \\
\hline                                                            
$0.15115   $ & ${\bf Old} $ & $5.79(48)$ & $-      $ & $-1.160(83)$ \\         
$0.2399(36)$ & ${\bf Exp} $ & $7.78(71)$ & $-      $ & $-1.48(11) $ \\         
             & ${\bf Lin} $ & $6.43(49)$ & $-      $ & $-1.267(81)$ \\         
             & ${\bf Sqr} $ & $7.7(14) $ & $2.6(28)$ & $-1.48(22) $ \\
\hline                                                                                             
$0.15100   $ & ${\bf Old} $ & $5.63(42)$ & $-      $ & $-1.115(70)$ \\        
$0.2924(36)$ & ${\bf Exp} $ & $7.42(60)$ & $-      $ & $-1.400(93)$ \\        
             & ${\bf Lin} $ & $6.19(42)$ & $-      $ & $-1.206(69)$ \\        
             & ${\bf Sqr} $ & $7.3(13) $ & $2.3(24)$ & $-1.39(19) $ \\
\hline                                                                                              
$0.15075   $ & ${\bf Old} $ & $5.33(36)$ & $-      $ & $-1.042(59)$ \\        
$0.3815(38)$ & ${\bf Exp} $ & $6.87(51)$ & $-      $ & $-1.282(76)$ \\        
             & ${\bf Lin} $ & $5.80(36)$ & $-      $ & $-1.118(58)$ \\        
             & ${\bf Sqr} $ & $6.5(11) $ & $1.5(21)$ & $-1.23(16) $ \\
\hline                                                                                              
$0.15050   $ & ${\bf Old} $ & $5.01(33)$ & $-      $ & $-0.973(54)$ \\        
$0.4728(40)$ & ${\bf Exp} $ & $6.34(45)$ & $-      $ & $-1.177(67)$ \\        
             & ${\bf Lin} $ & $5.42(33)$ & $-      $ & $-1.038(52)$ \\        
             & ${\bf Sqr} $ & $5.81(99)$ & $0.8(19)$ & $-1.10(15) $ \\
\hline                                                                                              
$0.15000   $ & ${\bf Old} $ & $4.36(30)$ & $-      $ & $-0.842(48)$ \\        
$0.6634(45)$ & ${\bf Exp} $ & $5.37(39)$ & $-      $ & $-0.996(58)$ \\        
             & ${\bf Lin} $ & $4.70(30)$ & $-      $ & $-0.894(46)$ \\        
             & ${\bf Sqr} $ & $4.72(89)$ & $0.0(17)$ & $-0.90(14) $ \\
\end{tabular}
\end{center}
\caption{\label{result.63.table}
The results at $\beta=6.3$.
Four lines for each $m_\pi$ are results with the fitting functions 
${\bf Old}$, ${\bf Exp}$, ${\bf Lin}$, and ${\bf Sqr}$,
which are defined in (\ref{eq:sqr})--(\ref{JLQCD.OLD_eq}).
}
\end{table}
%
%
\newpage
\begin{table}[t]
\begin{center}
\begin{tabular}{cc llll }
$\beta $ & $a\ (1/{\rm GeV})$ 
& ${\bf \quad Old}$ 
& ${\bf \quad Exp}$
& ${\bf \quad Lin}$
& ${\bf \quad Sqr}$ \\
\hline
$5.9$    & $0.493(7)$ & $-0.96(10) $ & $-1.51(16) $ & $-1.093(90)$ & $-1.58(36)$ \\
$6.1$    & $0.378(6)$ & $-1.185(59)$ & $-1.653(80)$ & $-1.335(55)$ & $-1.78(22)$ \\
$6.3$    & $0.302(5)$ & $-1.335(76)$ & $-1.745(99)$ & $-1.466(74)$ & $-1.77(21)$ \\
\hline                                                             
         & $a \to 0 $ & $-1.92(25) $ & $-2.09(35) $ & $-2.07(24) $ & $-2.04(78)$ \\
\end{tabular}
\end{center}
\caption{\label{a0-mp.C-limit.table}
The values of $a_0/m_\pi (1/{\rm GeV}^2)$ in the chiral limit
for the each fitting function of $R(t)$ at each $\beta$
and those in the continuum limit obtained by liner extrapolation in the lattice spacing.
The fitting functions of $R(t)$ are defined in (\ref{eq:sqr})--(\ref{JLQCD.OLD_eq}).
}
\end{table}
%
%
\end{document}